\documentclass[]{spie}  
\usepackage[]{graphicx}
\title{Exoplanet detection with simultaneous spectral differential imaging: effects of out-of-pupil-plane optical aberrations} 

\author{Christian Marois, Don W. Phillion \& Bruce Macintosh\supit{a}
\skiplinehalf
\supit{a}Institute of Geophysics and Planetary Physics L-413,\\ Lawrence Livermore National Laboratory, 7000 East Ave, Livermore, CA 94550, USA
}

\authorinfo{Send correspondence to C.M.: cmarois@igpp.ucllnl.org}

\begin{document} 
  \maketitle 

\begin{abstract}
Imaging faint companions (exoplanets and brown dwarfs) around nearby stars is currently limited by speckle noise. To efficiently attenuate this noise, a technique called simultaneous spectral differential imaging (SSDI) can be used. This technique consists of acquiring simultaneously images of the field of view in several adjacent narrow bands and in combining these images to suppress speckles. Simulations predict that SSDI can achieve, with the acquisition of three wavelengths, speckle noise attenuation of several thousands. These simulations are usually performed using the Fraunhofer approximation, i.e. considering that all aberrations are located in the pupil plane. We have performed wavefront propagation simulations to evaluate how out-of-pupil-plane aberrations affect SSDI speckle noise attenuation performance. The Talbot formalism is used to give a physical insight of the problem; results are confirmed using a proper wavefront propagation algorithm. We will show that near-focal-plane aberrations can significantly reduce SSDI speckle noise attenuation performance at several $\lambda/D$ separation. It is also shown that the Talbot effect correctly predicts the PSF chromaticity. Both differential atmospheric refraction effects and the use of a coronagraph will be discussed.
\end{abstract}



\section{INTRODUCTION}
\label{intro}  
Achieving high-contrast imaging on both ground- and space-based telescopes requires near perfect optics. Even with good optical surfaces, some quasi-static speckle noise remains that must be significantly attenuated to allow optimal photon noise limited sensitivity. Simultaneous spectral differential imaging (SSDI) \cite{racine1999,marois2000,sparks2002,biller2004,marois2004phd,marois2004,marois2005} is a technique that has the potential to significantly attenuates the speckle noise. SSDI consists of acquiring simultaneously in adjacent wavelengths multiple images of the field of view (FOV) and then combining them, after spatially scaling them by the ratio of the wavelength to correct for the speckle radial position dependence on wavelength, to suppress the speckle noise. Speckle noise is well correlated since images are acquired simultaneously. SSDI numerical simulations \cite{marois2000,marois2004phd} predict that speckle noise attenuation of several thousands is theoretically possible. If the wavelengths are chosen near a sharp spectral feature in the companion spectrum that is not present in that of the primary, speckle noise can be attenuated while retaining most of a companion's flux at all angular separations \cite{marois2004phd}.\footnote[1]{Even if the companion do not possess a spectral feature across the selected bands, it is still possible to detect it with SSDI due to the image magnification required to properly subtract the speckle noise, but in this case, the companion residual flux will show a positive and negative signal and it will be partially subtracted at small angular separations (see Ref.~\citenum{marois2004phd} for more details).} On-sky SSDI testing has shown that SSDI is currently limited by non-common path aberrations \cite{marois2005}. These non-common path aberrations produce quasi-static uncorrelated speckle noise in each optical channel \cite{marois2003,marois2005} that prevent an accurate speckle noise subtraction. New instrument designs, like the multi-color detector assembly \cite{marois2004} or similar concepts \cite{david2004,david2006}, have the potential to eliminate the non-common path aberration problem from multi-channel instruments by having a single optical channel for all wavelengths, and thus potentially achieving theoretical speckle noise attenuation predictions.

The usual technique to estimate theoretical SSDI speckle noise attenuation performance is to simply use the Fraunhofer approximation and perform the Fourier transform (FT) of a complex pupil. Phase aberrations (in radian) are scaled by the ratio of the wavelength to produce the chromatic PSF. Such approach is optimistic since all wavefronts at different wavelengths are perfectly aligned and all aberrations are located in the pupil plane.

In this paper, we analyze how out-of-pupil-plane phase aberrations affect SSDI performances by using a Talbot wavefront propagation software. Simulations include a typical differential atmospheric refraction effects and a coronagraph case is also studied. Results are confirmed using a fast Fourier transform-based code. Such analysis is fundamental for current high-contrast imaging projects, like the Gemini Planet Imager (GPI) and the VLT Planet Finder (VLTPF) as well as for future high-contrast space observatories. 
 \section{Out-of-Pupil-Plane Optical Aberrations\label{theo}}
\begin{figure}
   \begin{center}
   \begin{tabular}{c}
   \includegraphics[height=5cm]{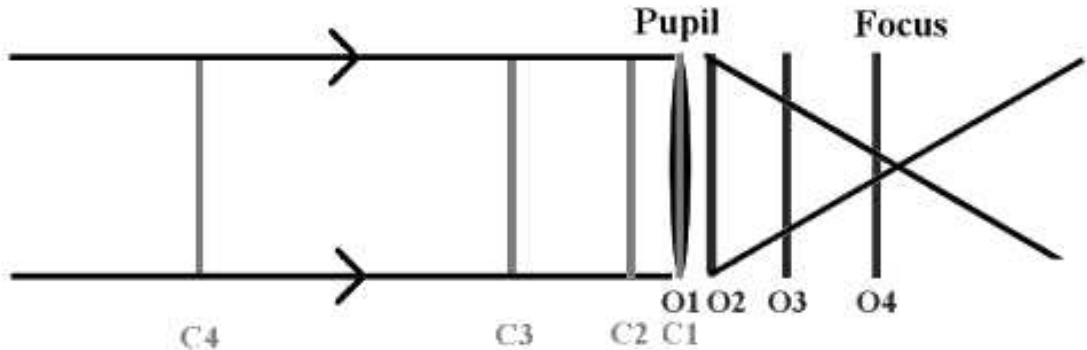}
   \end{tabular}
   \end{center}
   \caption[example] 
   {Simulation schematic. Aberrated surfaces are located at O1, O2, O3 and O4 (respectively at the pupil and at 10\%, 50\% and 90\% of the focal length) in between the pupil and focal planes. Corresponding conjugated plane in the collimated beam are found in front of the pupil (position C1, C2, C3 and C4). \label{fig1} }
   \end{figure}

For simplicity, we consider a simple optical system containing a collimated beam, a pupil, an aberrated optical surface and a focal plane. Aberrated optics are located in between the pupil and focal planes. PSFs are obtained by first finding the corresponding conjugate plane of the aberrated optic in the collimated beam above the pupil, by propagating the aberrated wavefront to the pupil plane and by Fourier transform of the complex pupil. As the aberrated optic gets closer to the focal plane, the corresponding conjugated plane in the collimated beam above the pupil goes to infinity (see Fig.~\ref{fig1}).

\noindent Conjugated wavefronts are propagated to the pupil by using Talbot imaging. Such an approach, though not exact for finite optics, is useful to get physical insight into wavefront propagation problems. Results obtained with this technique are of the same order of magnitude as the one obtained from a fast Fourier transform-based code for our simulated cases (see Sect.~\ref{simul}). Talbot imaging stipulates that when a wavefront propagates by a certain length, an initial phase aberration will oscillate between being a pure phase and a pure amplitude aberration. The length for a complete cycle is called the Talbot length $\tau_L$ \cite{talbot1836}. Propagating a phase aberration of a specific frequency by a quarter of its Talbot length will result in a pure amplitude aberration, while propagating by half the Talbot length will result in a pure phase aberration that has the opposite sign as the original phase aberration. Since the Talbot length is chromatic, the ratio of aberration in phase and amplitude space is not the same in two adjacent wavelengths, the two PSFs will be partially decorrelated and the SSDI speckle noise attenuation performance will not be as good as expected. The Talbot length is proportional to twice the ratio of the aberration spatial period $\Lambda$ squared over the wavelength $\lambda$

\begin{equation}
\tau_L = \frac{2\Lambda^2}{\lambda}.
\end{equation}

\noindent If the wavefront propagates by a length $S$ to reach the pupil plane, the total number of Talbot lengths traveled $N_{TL}$ is simply the ratio of $S$ over $\tau_{L}$

\begin{equation}
N_{TL} = \frac{S}{\tau_L}.
\end{equation}

\noindent The PSF chromaticity can be understood by comparing the number of Talbot length $\Delta N_{TL}$ of a specific spatial period at two wavelengths

\begin{equation} 
\Delta N_{TL} = \frac{S}{2\Lambda^2}\Delta \lambda.\label{ntl}
\end{equation}

\noindent As the aberration spatial frequency increases, the difference in Talbot length will result in an oscillation between (1) both aberrations being in phase or in amplitude and (2) one in phase and the other one in amplitude. If both aberrations are in phase or in amplitude, the predictions derived for a simple difference (SD, subtraction of two images at two different wavelengths) using the usual FT of the complex pupil are valid, while when one is in phase and the other one is in amplitude, the speckle attenuation can be severely limited. It is thus expected, from the Talbot approximation, that SSDI speckle noise attenuation performance will oscillate between being good and bad as a function of field angle. To better understand this effect, we can expand the complex pupil FT using a Taylor expansion \cite{bloemhof2001,siv2002,perrin2003,marois2004phd}. An amplitude error $\epsilon$ having a FT equal to $E$ produces a PSF intensity modulation $I$ given by

\begin{equation}
I(\eta,\xi) = I_0 + 2 \Re[p(p^*\star E^*)]+|p\star E|^2 \label{eq1}
\end{equation}

\noindent where $I_0$ is the unaberrated PSF, $p$ is the pupil FT, $\Re$ is the real part of the term and the symbol $\star$ is for a convolution; $\eta$ and $\xi$ are coordinates in the image plane. Both second and third aberration terms are symmetric since $E$ is Hermitian. A phase aberration $\phi$ having a FT equal to $\Phi$ modifies the PSF intensity as follow (truncated to the second order in $\Phi$)

\begin{equation}
I(\eta,\xi) \cong I_0 + 2\Im [p (p \star \Phi)] - \Re [p^* (p \star \Phi \star \Phi)]+|p \star \Phi|^2 \label{eq2}
\end{equation}

\noindent where the symbol $\Im$ is the imaginary part of the term. Since both phase and amplitude aberrations are real functions, the $\Re$ and $\Im$ parts are respectively symmetric and antisymmetric. For small aberrations and near diffraction rings, the second term of both equations dominates. If, at one wavelength, the aberration is in phase and in a second wavelength the aberration is in amplitude, subtraction of the PSF can be summarized, to first order, to the subtraction of a symmetric term to an antisymmetric term. This subtraction will clearly not work and important residuals will remain.

One interesting aspect of Eq.~\ref{ntl} is that it is linear in $\Delta \lambda$. If we select three wavelengths having equal spacing, the number of Talbot length $\Delta N_{TL}$ will be the same between the first and second wavelength and between the second and third wavelength. The two SDs ($I_{\lambda_2}$ - $I_{\lambda_1}$ and $I_{\lambda_2}$ - $I_{\lambda_3}$) would thus be correlated no matter how good the original SDs subtraction were, and a subtraction of these two SDs (called a double difference, or DD) will further attenuate speckles.

If a coronagraph is used, the second term of both Eq.~\ref{eq1} and Eq.~\ref{eq2} will be attenuated, leaving only symmetric terms. As mentioned by Ref.~\citenum{perrin2003} and numerically estimated by Ref.~\citenum{marois2004phd}, these symmetric terms can be removed by a 180 degree rotation and subtraction, leaving again the attenuated antisymmetric terms that have twice their original intensities. Again, the second term will be dominant and will not subtract well in a SD, but should be attenuated when performing a DD.

\section{Wavefront Propagation Simulations}\label{simul}
A Talbot wavefront propagation software was used to simulate the effect of an aberrated optical surface located at different position between a pupil and image planes. The simple and double differences are calculated by doing the propagation at three wavelengths. No 180 degree rotation and subtraction are performed here since it has been shown by Ref.~\citenum{marois2004phd}, for non-coronagraph cases, to only work well for very small aberrations; the gain is marginal for typical wavefront aberrations ($>$0.2 radian) and/or when a nearly equal mix of amplitude and phase errors are present.

For each simulation, aberrated optics are introduced (only one per simulation) at respectively the pupil and at 10\%, 50\% and 90\% of the focal length. Pupil size is 2~cm with a f/64 beam, chosen to be similar to the current Gemini Planet Imager (GPI) optical design. These simulations are thus aimed to better understand wavefront propagation inside the instrument, i.e. optics that are producing quasi-static speckles that are currently limiting exoplanet/brown dwarf detection. Corresponding conjugated planes above the pupil were found using the well known lens maker formula. The aberrated surfaces are produced using a power-law of index $-2.6$ having a conservative $50$~nm RMS of phase error (all surfaces are assumed to show the same quality over the beam size; expected phase errors for the GPI instrument is 5~nm rms per optic and 15~nm rms total, our simulations thus have approximately three times GPI expected wavefront total phase error). A wavefront tilt error was included to simulate a differential atmospheric refraction (DAR) between wavelengths (though this had minimal impact on the final speckle attenuation performance). Using the value found by the TRIDENT experiment at CFHT for 1.6 airmass (see Ref.~\citenum{marois2005}), a differential refraction of 10~mas (a quarter of a $\lambda/D$ at H-band on a 8-m telescope) is assumed between 1.515 and 1.625~$\mu $m. 

Each spatial frequency of the phase error (obtained by FT of the phase aberration) is Talbot propagated to the pupil by using Eq.~\ref{ntl} and split between a phase and amplitude aberrations depending on how many Talbot length occur to propagate to the pupil plane. Once all spatial frequencies have been propagated, i.e. the FT of the amplitude and phase aberrations are found, inverse FTs are performed to obtain the corresponding pupil plane amplitude and phase aberrations. This wavefront propagation is performed at three monochromatic wavelengths (1.515, 1.570 and 1.625$\mu $m). PSFs are simply obtained by FT of the complex pupil after properly normalizing the phase and amplitude aberrations as a function of wavelength. Images having $1024\times 1024$ pixels were used and pupil diameter was 256 pixels, producing PSF with 4 pixels per $\lambda/D$ (this choice produces PSFs that have the same number of pixels per $\lambda/D$ for all wavelengths, no spatial scaling is thus necessary to align speckles). Each PSFs are registered at the image center using an iterative cross-correlation algorithm.

\begin{figure}
   \begin{center}
   \begin{tabular}{c}
   \includegraphics[height=15cm]{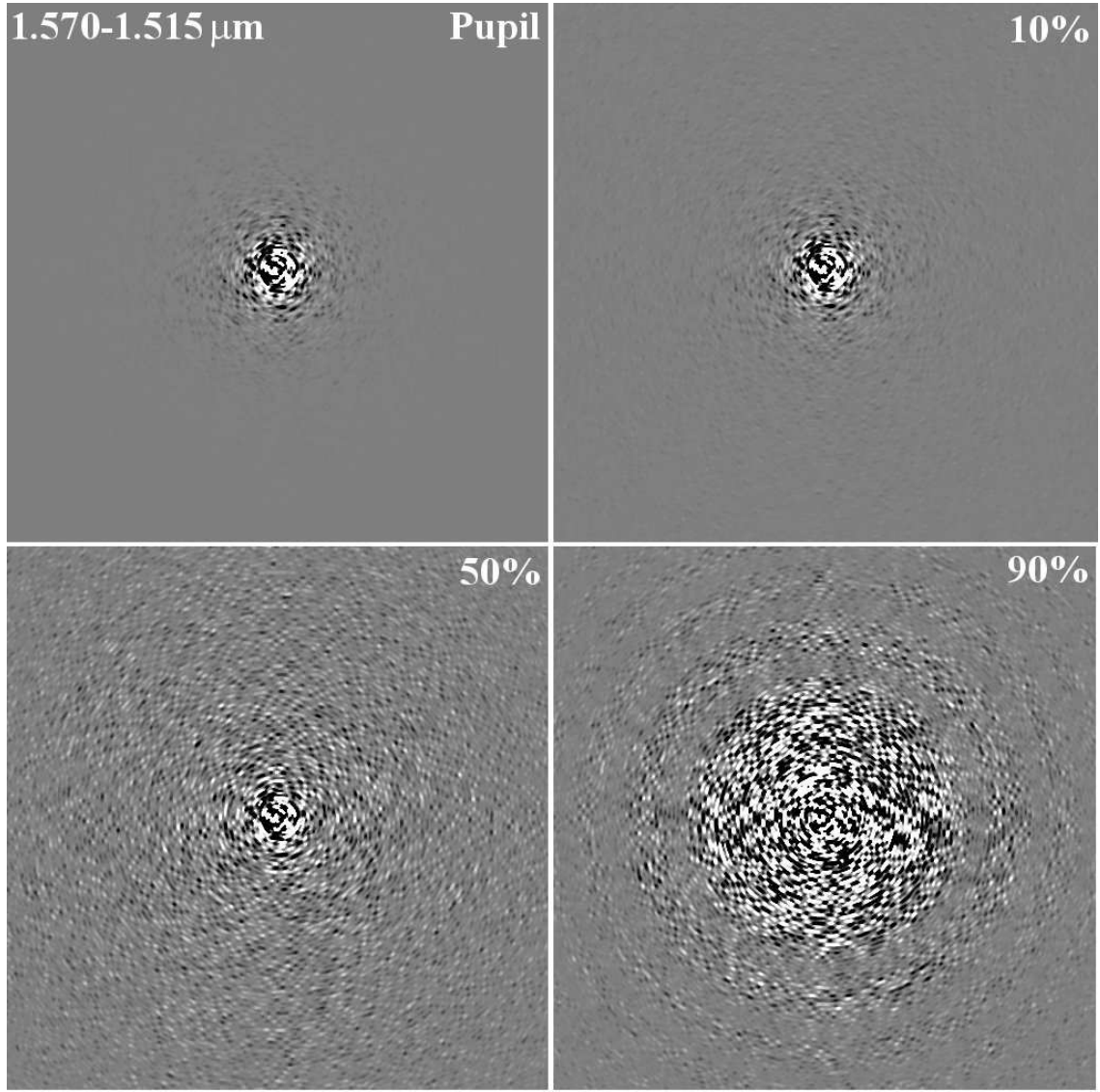}
   \end{tabular}
   \end{center}
   \caption[example] 
   {Simple difference (1.570 - 1.515$\mu $m) SSDI speckle noise attenuation performances with out-of-pupil-plane aberrations (50~nm rms). Aberrations are located at the pupil plane (upper left) and at respectively 10\% (upper right), 50\% (bottom left) and 90\% (bottom right) of the focal length. FOV is 128$\times $128 $\lambda/D$ and images are displayed with a linear gray scale between $\pm 2\times 10^{-6}$ of the PSF peak intensity. Wavefront beam has a 2~cm diameter and f/64 focal ratio.\label{fig1b}}
   \end{figure}

\begin{figure}
   \begin{center}
   \begin{tabular}{c}
   \includegraphics[height=11cm]{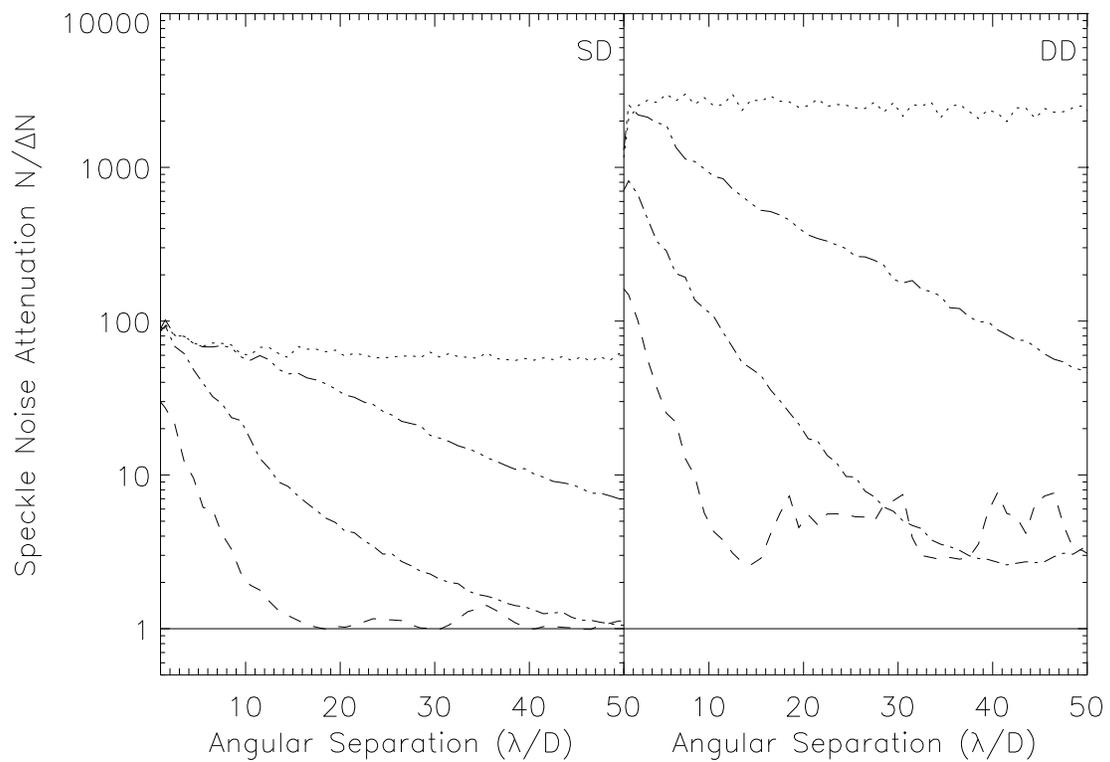}
   \end{tabular}
   \end{center}
   \caption[example] 
   {Talbot wavefront propagation. Left panel: speckle noise attenuation (N/$\Delta$N) for a SD obtained from the subtraction of two monochromatic PSFs at 1.515 and 1.570$\mu $m. Dotted line is for a pupil plane aberration (50~nm rms) and the triple-dot-dashed, dot-dashed and dashed lines are respectively for the same aberration located at 10\%, 50\% and 90\% of the focal length. Right panel: DD speckle noise attenuation obtained from the combination of three wavelengths at 1.515, 1.570 and 1.625$\mu $m. A 10~mas differential atmospheric refraction is included. Wavefront beam has a 2~cm diameter and f/64 focal ratio. \label{fig2}}
   \end{figure}

The speckle noise rms values are calculated for each angular separation inside a $\lambda/D$ width annulus. An aberration free PSF was generated for all non-coronographic cases and subtracted from each aberrated PSF before calculating SDs and the DD. Such step removes the unaberrated PSF term (see Eq.~\ref{eq1} and Eq.~\ref{eq2}) and leaves only the speckle noise, this noise can then be flux normalized at each wavelength to maximize speckle noise subtraction (see Ref.~\citenum{marois2004phd} for more details). The SDs and DD are simply obtained by the following two equations

\begin{equation}
SD_{j-i} = (I_{\lambda_j} - I_{\lambda_j}^0) - K_{SD_i} (I_{\lambda_i} - I_{\lambda_i}^0)
\end{equation}
\begin{equation}
DD = \frac{SD_{2-1} + K_{DD} SD_{2-3}}{2}.
\end{equation}

\noindent where $K_i$ are normalizing constants to optimize the speckle noise subtraction (the constant values are optimized for each annulus of 1~pixel width - similar equations have been found independently by Ref.~\citenum{deqing2006}\footnote[2]{Even if these algorithms show, in theory, better speckle attenuation performances than the original SSDI algorithms presented by Ref.~\citenum{marois2000}, in realistic conditions, it is difficult to imagine that any algorithm will do better than 100-1000$\times$ speckle attenuations due to systematic effects (i.e. flat field accuracy and primary/optical transmission spectrum differences between bandpasses) and non-common path aberrations \cite{marois2005} - for a 50~nm rms wavefront, speckle noise subtraction of more than 1 part by 1000 requires that non-common path aberrations, for non-coronagraphic images, be less than 0.05~nm RMS between channel, a nearly impossible task.}). The division by 2 in the DD is needed to insure that a companion has the same intensity as in the $I_{\lambda_2}$ image. Typical SD speckle noise attenuation performances are illustrated in Fig~\ref{fig1b}. Oscillations between good and bad speckle noise subtraction is clearly visible when the aberration is located 90\% of the focal length, as expected from the Talbot imaging theory \footnote[3]{The speckle subtraction performance oscillation is not visible when performing a complete optical wavefront propagation due to the finite aperture and diffraction effects; however, it will be shown that Talbot imaging properly predicts the expected PSF chromaticity}. The SD and DD noise attenuation $N/\Delta N$ is obtained by estimating the ratio of the initial PSF speckle noise $N$ in increasing annulus of $\lambda/D$ width over the one of the SD or DD $\Delta N$. The speckle noise attenuation for the simple and double differences are then smoothed by averaging the speckle noise attenuation of five independent simulations (see Fig.~\ref{fig2}). 

The simulation confirms that out-of-pupil-plane aberrations significantly influence the SSDI speckle noise attenuation performance, the effect becoming stronger as the aberrated plane is located closer to the focal plane as well as for bigger angular separations. For all cases, better speckle noise attenuation is achieved by the DD, as expected from the theory. For the simulation with the aberration located close to the focal plane, at small separations (10~$\lambda/D$), SD is already 50 times less precise than the case when the aberration is located at the pupil plane, while, for the DD, the noise is 1000 times worse than the pupil case. Both the SD and DD achieve speckle noise attenuation less than 10 at several $\lambda/D$ separation.

\begin{figure}
   \begin{center}
   \begin{tabular}{c}
   \includegraphics[height=11cm]{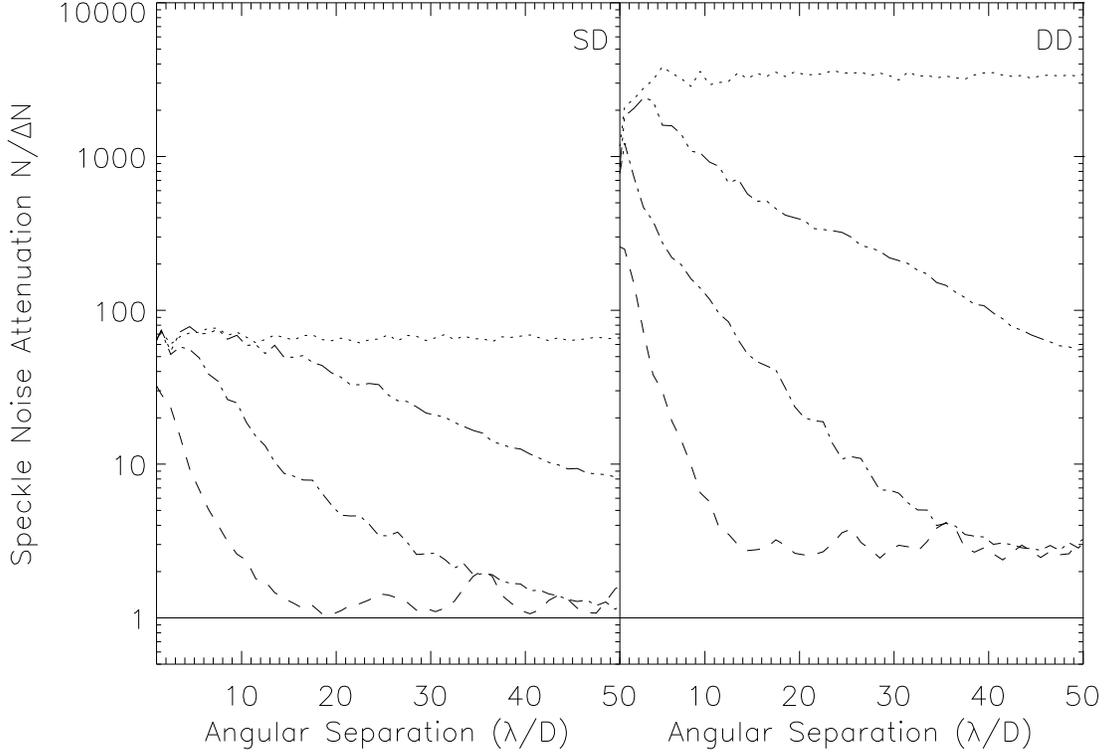}
   \end{tabular}
   \end{center}
   \caption[example] 
   {Talbot wavefront propagation. Same as Fig.~\ref{fig2}. A coronagraph, simulated using a Gaussian apodizer, is included. \label{fig5}}
   \end{figure}

\begin{figure}
   \begin{center}
   \begin{tabular}{c}
   \includegraphics[height=11cm]{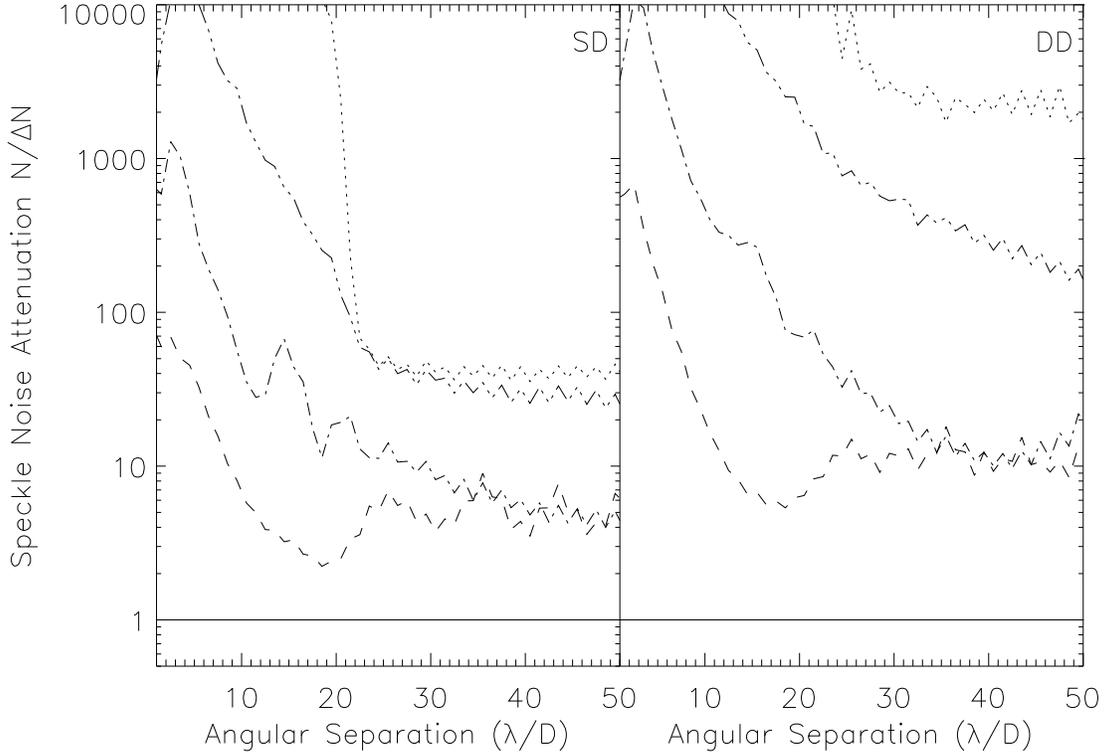}
   \end{tabular}
   \end{center}
   \caption[example] 
   {Talbot wavefront propagation. Same as Fig.~\ref{fig2}. A coronagraph, simulated using a Gaussian apodizer, and a H-band pupil plane perfect phase correction (inside the SFWFS control radius) are included.\label{fig6}}
   \end{figure}

A coronagraph was also simulated using a Gaussian apodizer (the pupil was convolved by a pupil having one third of the pupil diameter). A 10~mas DAR effect was included between 1.515 and 1.625$\mu $m. Two simulations, with and without a H-band pupil plane phase correction using an ideal spatial filtered wavefront sensor (SFWFS) \cite{poyneer2004}, i.e. aberrations are attenuated to zero inside the deformable mirror control radius (half the actuator spatial frequency - a $40\times 40$ actuators deformable mirror is assumed), were also performed to simulate the GPI precision wavefront sensor \cite{wallace2004}. These corrections are performed at 1.65~$\mu $m, the average wavelength of the broad band H filter. The H-band pupil phase aberration is obtained, like the other three wavelenths, by Talbot propagating the conjugated aberration. Fig.~\ref{fig5} and \ref{fig6} shows the SD and DD speckle noise attenuation for these two cases.

\noindent For an aberrated optic located near the focal plane, the speckle noise attenuation obtained with a coronagragh is similar to the non-coronagraph case. The coronagraph thus does not degrade, in theory, achievable speckle noise attenuation. For optics located near the pupil plane, the precise H-band phase correction gain is impressive ($>$10), but shows limited gain ($\sim 2$) for near-focal-plane optics.

\begin{figure}
   \begin{center}
   \begin{tabular}{c}
   \includegraphics[height=11cm]{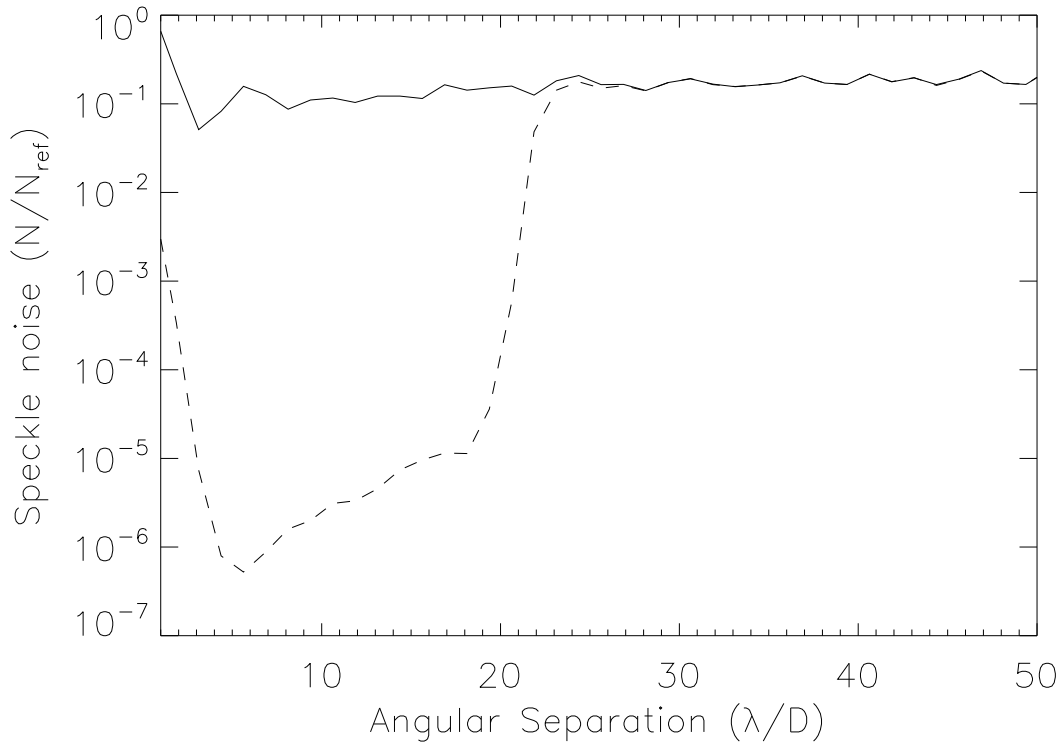}
   \end{tabular}
   \end{center}
   \caption[example] 
   {Relative speckle noise (from pupil plane aberration) for the three simulated cases. The reference speckle noise $N_{\rm{ref}}$ is the one from the phase only simulation (see Fig.~\ref{fig2}). The solid line is for the coronagraph simulation (see Fig.~\ref{fig5}) and the dashed line is for the coronagraph with pupil plane phase correction simulation (see Fig.~\ref{fig6}).\label{fig7}}
   \end{figure}

The last three simulations showed that SSDI speckle noise attenuation is essentially unaffected by the coronagraph and that a precise pupil plane phase correction can increase performances at small offset only if aberrations are located near the pupil plane. These performances, when aberrations are located near the pupil plane, can be explained by the PSF speckle noise as a function of angular separation. Fig~\ref{fig7} shows the relative speckle noise, from pupil plane aberration, of the non-coronagraph case compared to the speckle noise of the last two simulations (with coronagraph and with/without pupil plane phase correction). The coronagraph reduces the speckle noise by a factor $\sim 5$ at all angular separations while the pupil plane phase correction greatly reduces the speckle noise for separation less than 20~$\lambda/D$. As it was discovered by Ref.~\citenum{marois2004phd}, SSDI (SD and DD) speckle noise attenuation performances with adequate speckle noise normalization are greatly increased as aberrations get smaller. This is mainly due to the fact that the Taylor expansion converges more rapidly for smaller aberrations (ratio of terms of following orders get bigger as the aberration get smaller - better speckle noise attenuation is thus obtained when subtracting 1 (SD) or 2 (DD) orders using SSDI).

\begin{figure}
   \begin{center}
   \begin{tabular}{c}
   \includegraphics[height=11cm]{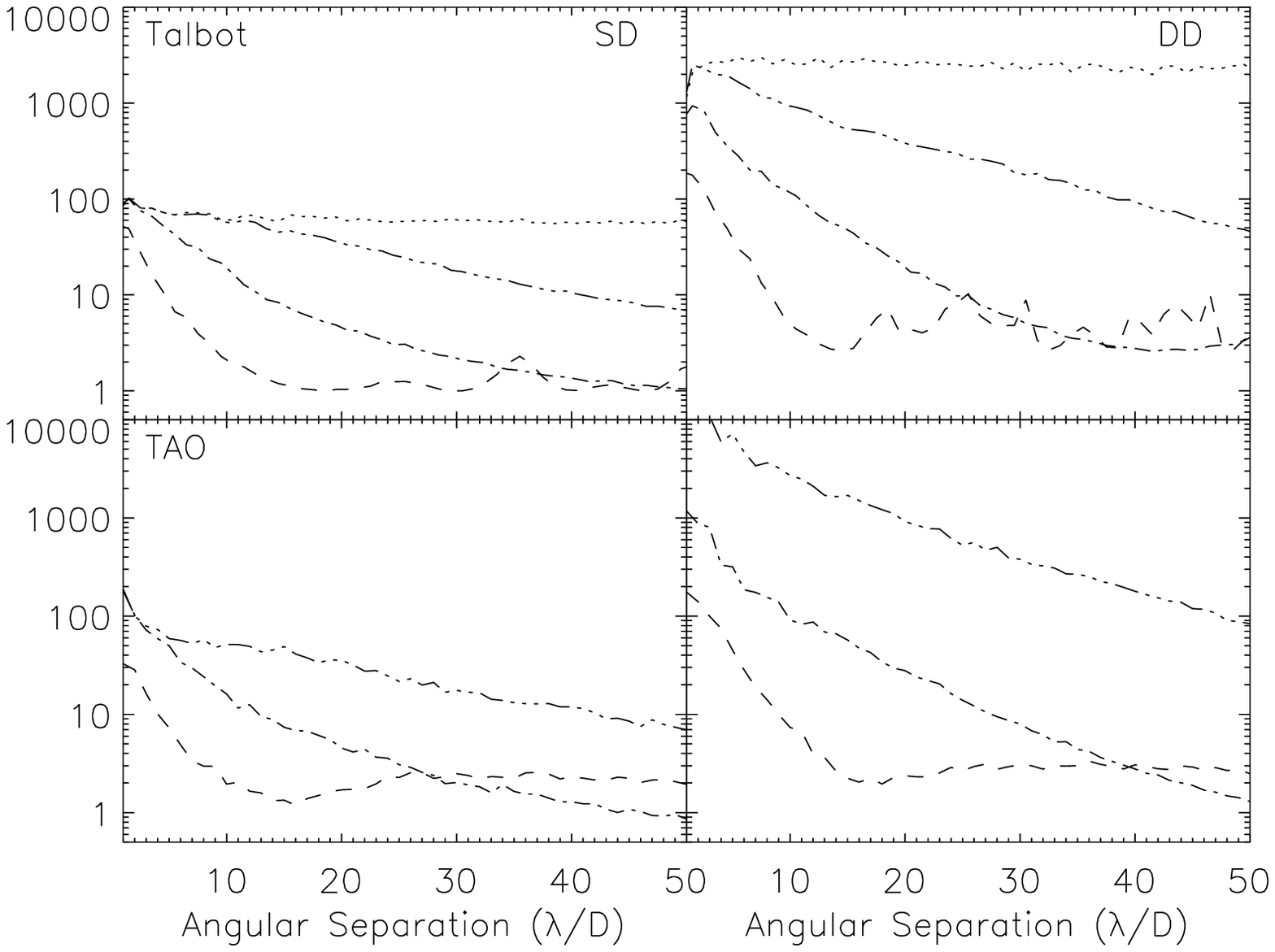}
   \end{tabular}
   \end{center}
   \caption[example] 
   {Talbot imaging compared to a FFT-based optical wavefront propagation code. The upper two panels show the Talbot imaging SD and DD results (same as Fig.~\ref{fig2}) while the two bottom panels show the same SSDI simulation performed with a FFT-based wavefront propagation code (no simulation was performed for aberrations located at the pupil plane).\label{fig8}}
   \end{figure}

Talbot imaging is a very simple propagation algorithm, other known effects, e.g. the finite aperture, are not simulated. The overall SSDI speckle attenuation performance is thus expected to be worse than the one presented in this paper when doing full optical propagation. To verify the validity of our Talbot imaging simulations, we have performed a FFT-based simulation for the non-coronagraph case (see Fig.~\ref{fig8}). The FFT-based code is called the Telescope AO code (TAO).\footnote[4]{The TAO code can be obtained by email from phillion1@llnl.gov and documentations can be found at:  http://library.llnl.gov/uhtbin/cgisirsi/0/0/0/60/55/X.} This code does numerical propagation using the complex FFT and makes the paraxial approximation. It is shown that Talbot propagation actually fit very well the TAO simulations. It this thus tempting to conclude that the Talbot effect, for out-of-pupil-plane aberrations, is what is mostly limiting SSDI image subtraction.

\section{Discussion}
Simulations presented in this paper assumed a single aberrated surface for simplicity. If there are multiple aberrated surfaces located at different places, speckle suppression can be further limited if these surfaces are sufficiently apart from each other in the collimated beam (see Fig.~\ref{fig1}). Here is an example of how speckle suppression can be affected if they are too far apart: Suppose there are two surfaces in the collimated beam which introduce exactly the same sinusoidal phase perturbation. Suppose that at the first wavelength, the spacing between them is an even multiple of half the Talbot length while at second wavelength, the spacing between them is an odd multiple of half the Talbot length. Then at the first wavelength, the sinusoidal phase fluctuations add and the speckle amplitudes double, while at the second wavelength, the sinusoidal phase fluctuations exactly cancel and there are no speckles. Clearly, SSDI speckle noise attenuation cannot be done at this spatial frequency. What is too far apart depends upon the aberration spatial period. For sufficiently large spatial period $\Lambda$, the relayed aberrated surfaces will be sufficiently close together compared to the Talbot length to not further degrade speckle suppression performance.

Although Talbot imaging cannot in general be used for more than one aberrated surface, the Talbot length itself is still useful even when there are multiple aberrated surfaces. Talbot imaging assumes that for each sinusoidal fluctuation, there is a plane in which that fluctuation is purely in phase. However, it is possible to create a speckle for which has no opposing speckle when there are multiple aberrated surfaces. Here's an example of how this can happen: Suppose there are two surfaces spaced a quarter Talbot length apart in the collimated beam and that on the first surface there is a sine wave phase perturbation and that on the second surface there is a cosine wave phase perturbation which has the same magnitude and same period. The sine wave phase fluctuation from the first surface becomes a sine wave amplitude fluctuation when propagated to the second surface.  After the cosine wave phase fluctuation from the second surface is added, one speckle cancels out and its opposing speckle doubles in amplitude. When there is only one speckle, it creates a running wave in which the phase and amplitude fluctuations are ninety degrees out of phase and in which the peak phase fluctuation in radians equals the peak fractional amplitude fluctuation.  Regardless of whether one has a pure standing wave caused by two equal magnitude speckles, or has a pure running wave caused by a single speckle, or has some mixture of the two, the Talbot length is still the length for which the complex wavefront repeats. In the first case, the standing wave rotates in phase-amplitude space along the propagation axis while in the second case the phase-amplitude pattern moves transversely.

The Talbot effect also has important implications for speckle symmetry. It is expected, from the Fraunhofer approximation, that high Strehl ratio non-coronagraphic PSFs and phase only aberrations will show anti-symetries due to the dominant anti-symetric first speckle term of Eq.~\ref{eq2}.\cite{bloemhof2001} If out-of-pupil-plane aberrations are present, some phase aberrations at certain spatial frequencies will become amplitude aberrations. These amplitude aberrations are producing a symmetric pin term (see Eq.~\ref{eq1}) that can be of the same order of magnitude as the anti-symmetric term, resulting in a high Strehl ratio PSF with a speckle noise not showing any dominant symmetry.

In a real instrument design, a proper wavefront propagation tool is needed to correctly simulate the PSF (Fresnel, FFT, Rayleigh-Sommerfeld, ...). Such algorithm will be used for a complete end-to-end simulations of GPI optical design to correctly specify the optical quality needed for each optical component to insure that SSDI speckle attenuation will reach the photon noise sensitivity limit.

\section{Conclusions}
Talbot wavefront propagations were used to show that out-of-pupil-plane aberrated optics can significantly limit SSDI speckle noise attenuation performances, especially for wide separations and/or if the aberrated optic is located near a focal plane. SDs and DD can be 50 to 1000 noisier than expected from simple Fraunhofer simulations for separations greater than 10~$\lambda/D$. If a calibration system working at the science wavelength is used, it was shown that relatively good speckle attenuation can be obtained at all separations inside the AO dark hole if the aberrated surface is located near the pupil plane. If several optics are located near a focal plane, such correction offers limited gain in SSDI speckle noise attenuation performances, even with a coronagraph.

It was possible to subtract the speckle noise using three adjacent wavelengths by more than a factor $>10$ (up to several thousands) using a DD and for separations less than 5~$\lambda/D$ (0.2$^{\prime \prime}$\ for H-band imaging on a 8-10~m telescope). SSDI can help reduce the speckle noise by more than a factor of $\sim $10 inside 10~$\lambda/D$, but SSDI will not achieved the predicted $>100$ speckle noise attenuation as derived by the Fraunhofer approximation at all angular separations. Near-focal-plane optics must be avoided or carefully specified if SSDI is to be used for $>10-100\times$ speckle noise attenuation.

Talbot imaging correctly predicts the PSF chromaticity. It is tempting to conclude that the Talbot chromatic phase to amplitude oscillation is the dominant effect that prevents accurate SSDI speckle noise attenuation for out-of-pupil-plane optics in the regime simulated in this paper.

Out-of-pupil-plane optics, along with non-common path aberrations in multi-channel instruments, are probably what is currently limiting existing SSDI instruments, like TRIDENT and the Simultaneous Differential Imager (SDI) at VLT.

\acknowledgments     
 
This research was performed under the auspices of the US Department of Energy by the University of California, Lawrence Livermore National Laboratory under contract W-7405-ENG-48, and also supported in part by the National Science Foundation Science and Technology Center for Adaptive Optics, managed by the University of California at Santa Cruz under cooperative agreement AST 98-76783.


\end{document}